**Title**

# Unraveling the orientation of phosphors doped in organic semiconducting layers

**Authors**

Chang-Ki Moon, Kwon-Hyeon Kim, and Jang-Joo Kim*

Department of Materials Science and Engineering, RIAM, Seoul National University, Seoul 151-744, South Korea.

E-mail: jjkim@snu.ac.kr

**Abstract**

Emitting dipole orientation (EDO) is an important issue of emitting materials in organic light-emitting diodes (OLEDs) for an increase of outcoupling efficiency of light. The origin of preferred orientation of emitting dipole of spherically shaped iridium-based heteroleptic phosphorescent dyes doped in organic layers is revealed by simulation of vacuum deposition using molecular dynamics (MD) along with quantum mechanical (QM) characterization of the phosphors for a direct comparison with experimental observations of EDO. Consideration of both the electronic transitions in a molecular frame and the orientation of the molecules interacting with the environment at the vacuum/molecular film interface allows quantitative analyses of the EDO depending on host molecules and dopant structures. Interaction between the phosphor and nearest host molecules on the surface, minimizing the non-bonding energy determines the molecular alignment during the vacuum deposition. Parallel alignment of the main cyclometalating ligands in the molecular complex due to host interactions rather than the ancillary ligand orienting to vacuum leads to the horizontal EDO.

Orientation of molecules in molecular films dictates their electrical and optical properties such as charge mobility,[1-2] birefringence,[3] absorption,[4] emission,[5] ionization potential,[6] and dielectric[7] and ferroelectric properties.[8] Therefore, understanding and control of molecular orientation in organic films have been a research topic with central importance in organic electronics and photonics, including the fields of liquid crystals,[9] organic field effect transistors,[10] and organic photovoltaics.[11] In organic light-emitting diodes (OLEDs), the molecular orientation of emitter embedded in the emissive layer has been an issue to enhance the outcoupling efficiency of light pursuing the horizontal alignment of the emitting dipole moment.[3,12-21]

Interestingly enough, it is only in recent years has attention turned to the orientation of emitting dipoles of iridium-based phosphors, the most verified light emitting dyes with high photoluminescence (PL) quantum yield and variety of chromatic spectrum as doped in the emissive layers; probably because their iridium-centered spherical shape and the amorphous surrounding nature in the emissive layers are far from having strong molecular alignments. Recently, some heteroleptic Ir complexes exhibiting efficient electroluminescence in OLEDs are reported to possess preferred horizontal emitting dipole orientations (EDOs).[13-16,18-20] However, it was difficult to assert the reason why the spherical-shaped phosphors have a propensity toward preferred molecular alignment in the emissive layers. A few mechanisms have been proposed to explain the preferred molecular orientation of the Ir complexes doped in vacuum deposited organic semiconducting layers: (1) molecular aggregation of the dopants leading to randomizing their orientation by suppressing the intermolecular interaction between the dopant and host molecules,[22] (2) strong intermolecular interactions between electro-positive sides of the dopant and the electro-negative host molecules promoting parallel alignment of the *N*-heterocycles of Ir-complexes by forming host-dopant-host pseudo-complex mainly participating in $^3$MLCT transition,[16,23] and (3) π-π interactions between the dopant and host molecules on the organic surface bringing alignment of aliphatic ligands to the vacuum side.[20,24] Currently, it is not very clear which mechanism most comprehensively describes the origin of the preferred EDO of the heteroleptic iridium phosphors. Moreover, the models are too oversimplified to describe the EDO values quantitatively, which depend on structures of the phosphors and host molecules.[23] Therefore, the molecular configurations and the interactions responsible for the EDOs of Ir-complexes should be established by atomic-scale simulation of the Ir complexes interacting with host molecules during film fabrication.

In this paper, we carefully examine the vacuum deposition process of phosphors on organic layers using a combination of molecular dynamics (MD) simulations and quantum mechanical (QM) analyses. The triplet EDO of heteroleptic Ir complexes doped in organic layers are studied with systematic variations of the molecular structures of both host and dopant. Theoretical prediction of EDO from simulated deposition process reveals excellent quantitative agreement with experimental observations, reproducing the anisotropic molecular orientations of heteroleptic Ir complexes in the emissive layers. In-depth analysis indicates that the molecular orientation originates from the coupling of the cyclometalated main ligand participating in the optical transition with neighbor host molecules rather than from the alignment of aliphatic ancillary ligand toward vacuum. Close observation of the simulation results indicate that non-bonded energy has a critical influence on the molecular orientation during the deposition.

## Results

### *Modeling of emitting dipole orientation*

The simulation method for obtaining EDO of an emitter in the vacuum deposited layer is schematically illustrated in Fig. 1a. Firstly, the transition dipole moment (TDM) vector in the molecular frame ($\boldsymbol{m}_X$-, $\boldsymbol{m}_Y$-, and $\boldsymbol{m}_Z$-axes) was determined by QM calculations after optimization of molecular geometry. For iridium-based phosphors, spin-orbit coupled time-dependent density functional theory (SOC-TDDFT) was employed for the calculation of the triplet TDM vectors for phosphorescence. Secondly, vacuum deposition of the emitting molecules on organic surfaces was simulated using MD. Finally, the TDM vectors in the molecular axis in each frame of MD were transformed to the vectors in the laboratory axis ($\boldsymbol{n}_X$-, $\boldsymbol{n}_Y$-, and $\boldsymbol{n}_Z$-axes) by rotation matrix method (Fig. 1b). We determine $\varphi_C$ and $\varphi_L$ as the angle between $\boldsymbol{m}_Z$ and $\boldsymbol{n}_Z$ axes and the angle between the TDM vector of the emitter and $\boldsymbol{n}_Z$ axis, representing the molecular orientation and the EDO against the vertical direction in the laboratory axis, respectively. The ratio of the horizontal $(\mathrm{TDM_H})$ to the vertical transition dipole moment $(\mathrm{TDM_V})$ follows the trigonometric relationship:

$$\mathrm{TDM_H} : \mathrm{TDM_V} = \mu_0^2 \sin^2 \varphi_L : \mu_0^2 \cos^2 \varphi_L, \tag{1}$$

where $\mu_0$ is the magnitude of the dipole moment and squares of the components indicate the intensity of the transition (emission intensity). The EDO describes an average fraction of the horizontal and vertical dipole moment of whole emitters embedded in the emissive layer. An ensemble average of the horizontal dipole moment gives the fraction of horizontal emitting dipole moment in the emissive layer (Θ) as a parameter of the EDO by

$$\Theta = \left\langle \sin^2 \varphi_{\mathrm{L}} \right\rangle . \quad (2)$$

Details about the rotation matrix and the vector transformation are given in Methods section.

The deposition simulation was performed by dropping a target molecule onto organic substrates under vacuum followed by thermal equilibration at 300 K as shown in Fig. 1c. The simulations were performed using the Materials Science Suite (Version 2.2) released by Schrödinger Inc.[25] Force field of OPLS_2005[26] and periodic boundary conditions were used for the MD simulations. Organic substrates were prepared by packing of 256 host molecules at 300 K and 1 atm. The simulated substrates have random molecular orientations and similar densities as the experimental results. Detailed steps of preparation of the substrates are given in Fig. S1 in Supplementary Information. One of the challenges of a single-trajectory-based MD analysis for orientation during deposition is that the time scale needed to observe the entirety of lateral degrees of freedom for a single molecule is much longer than that of a typical MD simulation. As such, we introduced 50 independent deposition events per dopant, instead of relying upon a single MD trajectory for each. 50 target dopant molecules were distributed in the vacuum slab of the periodic substrate model at un-overlapped locations with different orientations for the deposition simulation. Each target molecule was individually dropped onto the substrate under vacuum at 300 K. Translational motion of the host molecules at the bottom of the substrate was restrained in order to avoid the drift of the system. The deposition simulation used an NVT ensemble for a duration of 6000 ps with a time step of 2 fs and configurations of the system were recorded every 6 ps. One example of the process is shown in Supplementary video. Finally, EDOs of the phosphors and the molecular angles ( $\varphi_C$ ) were analyzed using equation (1) from the configurations. The analysis is based upon an assumption that the characteristic time to determine the orientation of dopants is in same scale of which the intermolecular interaction converges after the deposition of a dopant.

*Materials*

Chemical structures of the materials used in this study are depicted in Fig. 2a and 2b. Three heteroleptic iridium complexes of $Ir(ppy)_2tmd$, Ir(3′,5′,4-mppy)$_2$tmd,[18] and Ir(dmppy-ph)$_2$tmd[19] possessing high Θ values were adopted to investigate the effect of the phosphor molecular structure. The molecular $C_2$ symmetry axis toward the center of the ancillary ligand from the origin located at the Ir atom was set as $\boldsymbol{m}_z$, the orthogonal vector to $\boldsymbol{m}_z$ normal to the molecular Ir-O-O plane was set as $\boldsymbol{m}_x$, and $\boldsymbol{m}_y$ was determined by a cross product of $\boldsymbol{m}_z$ and $\boldsymbol{m}_x$ in the dopants. The triplet TDM vectors of the three Ir-complexes align along the direction of the iridium atom to the pyridine rings by $^3$MLCT as displayed in Fig. 2a. Coordinates of the TDM vectors of $Ir(ppy)_2tmd$, Ir(3′,5′,4-mppy)$_2$tmd, and Ir(dmppy-ph)$_2$tmd were [$\varphi_M$ = 88°, $\theta$ = 147°], [$\varphi_M$ = 89°, $\theta$ = 141°], and [$\varphi_M$ = 89°, $\theta$ = 156°], respectively, indicating that the substituents at the 4-position of the pyridine of the main ligands do not change the direction of triplet TDM vectors much.

Diphenyl-4-triphenylsilyphenyl-phosphineoxide (TSPO1), 1,4-bis(triphenylsilyl)benzene (UGH-2), and 4,4′-bis(*N*-carbazolyl)-1,1′-biphenyl (CBP) were selected as host materials to investigate the effect of ground state dipole and conjugation length of host on the EDO. TSPO1 has large permanent dipole moment due to the polar phosphine oxide group and the asymmetric structure, while UGH-2 and CBP molecules have small ground state dipole moments compared to TSPO1 due to the symmetric structures and the less polar groups. On the other hand, CBP has longer conjugation length than UGH-2 and TSPO1, indicating that CBP has larger polarizability than UGH-2.

Experimentally, Ir(ppy)$_2$tmd exhibited the Θ values of 0.60, 0.75 and 0.78 when doped in the UGH-2, CBP, and TSPO1 layers, respectively. Ir(3′,5′,4-mppy)$_2$tmd and Ir(dmppy-ph)$_2$tmd doped in TSPO1 layers have enhanced horizontal dipole orientation with the Θ values of 0.80 and 0.86, respectively (Fig. S2 in Supplementary Information and ref. 23).

*Simulation results*

The simulated variations of the orientation of the TDM$_H$ and $C_2$ axis of the dopants with time on the different hosts are displayed in Fig.S3 in Supplementary Information for 50 depositions for each system. The orientation of the phosphors was stabilized after certain time for some molecules, but fluctuated continuously for other molecules.

Fig. 3a exhibits the histograms of the EDO resulting from the deposition simulation. The blue lines represent the probability density of $\mathrm{TDM_H}$ ($\sin^2 \varphi_L$, Appendix 1) of an arbitrary vector. The green lines exhibit the deviations of the population from the random distribution. The simulated Θ values of Ir(ppy)$_2$tmd were 0.63, 0.72 and 0.74 on the UGH-2, CBP and TSPO1 substrates, respectively. Ir(3′,5′,4-mppy)$_2$tmd and Ir(dmppy-ph)$_2$tmd on TSPO1 substrates have the Θ values of 0.76 and 0.82, respectively. In addition, the simulation was performed for Ir(ppy)$_3$, a homoleptic complex exhibiting isotropic EDO when doped in CBP as a reference.[14,23] The distribution of the emitting dipole moment of Ir(ppy)$_3$ was close to the random distribution with a simulated Θ value of 0.67 and random orientation of the $C_3$ symmetry axis of the molecule (see Fig. S4 in Supplementary Information). The simulated EDOs match well with the experimental results as compared in Table 1, verifying that the MD simulation describes the vacuum deposition adequately. The results show that Ir(ppy)$_2$tmd in UGH-2 has larger molecular population with vertical TDM at the expense of reduced population with horizontal TDM compared to the random distribution (Θ=0.67). Higher Θ values are obtained when population of molecules possessing high TDM$_H$ is getting larger with the reduced population with low TDM$_H$.

The orientation of the $C_2$ axes of the phosphors on the organic layers are shown in Fig. 3b to find out if alignments of the aliphatic ancillary ligands have any correlation with EDO, for instance, if the horizontal EDO results from the vertical alignment of ancillary ligands with respect to the substrate.[20,24] One expects that the distribution function follows $\sin\varphi_c$ (blue line) if the orientation is random. We can extract several interesting results from Fig. 3b: (1) The orientation of the ancillary ligand of the Ir-complexes has broad distributions for all the deposited films. (2) Host effect on EDO is independent of alignments of the ancillary ligand. The total distribution of the $C_2$ axis of Ir(ppy)$_2$tmd are similar on the UGH-2, CBP, and TSPO1 layers with the average $\varphi_C$ of 70°, but the EDO on UGH-2 host is different from the EDOs on other two hosts. (3) The orientation of the $C_2$ axes of Ir(3,5′,4-mppy)$_2$tmd and Ir(dmppy-ph)$_2$tmd on TSPO1 are more random (closer to $\sin\varphi_c$) even though they possess higher Θ values than Ir(ppy)$_2$tmd. The random distributions are observed even in the region with high horizontal alignment of the emitting dipole moment (green regions in the stacked histogram with 0.95≤TDM$_H$≤1). (4) The dopant molecules with vertical TDM (red regions in the stacked histogram with 0≤TDM$_H$≤0.3) have $\varphi_C$ close to 90° for all the system, indicating that the ancillary ligands align parallel to the surface. All the results show that there is little correlation between the orientation of TDMs and alignment of the ancillary ligands.

## Discussion

We performed the MD simulation by depositing each target molecule on an organic substrate at a time and obtained statistics by repeating for 50 molecules deposited on different positions of the organic substrates so that the aggregation effect of the Ir complexes is neglected. Very good consistency of the simulated EDO and experimental values clearly indicates that aggregation is not a necessary condition for the alignment of heteroleptic Ir complexes.

Alignment of aliphatic ligands of heteroleptic Ir complexes to vacuum (model 3) is not required for preferred horizontal EDO either as shown in Fig. 3. Much larger portion of the aliphatic ligand (–tmd group) of $Ir(ppy)_2tmd$ molecules align to the vacuum side ($0° < \varphi_C < 90°$ in Fig. 3b) than Ir(3′,5′,4-mppy2tmd and $Ir(dmppy\text{-}ph)_2tmd$ molecules. However, the Θ value of $Ir(ppy)_2tmd$ is much lower than $Ir(3′,5′,4\text{-}mppy)_2tmd$ and $Ir(dmppy\text{-}ph)_2tmd$. These results are the reverse direction from the prediction based on the model and clearly demonstrate, therefore, that alignment of aliphatic ligands to the vacuum side is not a necessary condition for the alignment of EDO in heteroleptic Ir complexes. The reason why it is not required can understood from the following consideration.

The relationship between orientation of molecules and emitting dipole moment can be easily figured out using schematic molecular orientations of a heteroleptic Ir complex shown in Fig. 4a. The $C_2$ axis is toward the ancillary ligand (dark blue arrows) and the TDM vector (red arrows) is approximately along the direction from the iridium center to one of the pyridine rings. The alignment of iridium-pyridines determines the orientations of TDM for $Ir(ppy)_2tmd$, $Ir(3',5',4\text{-}mppy)_2tmd$, and $Ir(dmppy\text{-}ph)_2tmd$. Fig. 4a shows 5 configurations with different rotation angles of the $C_2$ axis for the horizontal TDM and 1 configuration for the vertical TDM. Rotation of the $C_2$ axis from the vertical to the horizontal direction can result in the horizontal TDM as long as the TDM is located on the horizontal plane (substrate) with an arbitrary orientation of the ancillary ligand. In other words, horizontal EDO is possible no matter which direction of the ancillary ligand aligns, either toward vacuum or film. On the other hand, the vertical TDM is obtained only when the pyridine rings are aligned perpendicular to the substrate. It accompanies horizontal alignment of the $C_2$ axis $(\varphi_C \sim 90°)$ on the configuration. This consideration is consistent with the simulation results in Fig. 3.

Non-bonding energy was calculated from the MD simulation to investigate if the intermolecular interaction between the phosphor and neighbor host molecules is responsible for the spontaneous molecular alignments of the phosphors on the surfaces. Fig. 4b depicts the correlation between non-bonding energy and orientation of the emitting dipole moment of the phosphors in the five different host-dopant systems. Distributions of the non-bonding energy are given in Fig S5 in Supplementary Information. There is a broad energy trap of ~ 3 kcal/mol in the region of $TDM_H$=0.1−0.5 for $Ir(ppy)_2tmd$ on the UGH-2 host and the energy increases with further increasing of $TDM_H$, thereby resulting in rather vertical EDO compared to random orientation because the population of $TDM_H$ is expected to be concentrated in the regions of low (large) non-bonding energy. On the other hand, non-bonding energies of $Ir(ppy)_2tmd$ on CBP and TSPO1 layers and the energies of $Ir(3',5',4\text{-}mppy)_2tmd$ and $Ir(dmppy\text{-}ph)_2tmd$ on TSPO1 are lowered as $TDM_H$ increases. As a result, molecular alignment with horizontal TDM is energetically preferred when they are deposited onto the organic semiconducting layers. Furthermore, much lower energies were obtained from Ir(3′,5′,4-mppy)$_2$tmd and Ir(dmppy-ph)$_2$tmd than Ir(ppy)$_2$tmd on the TSPO1 layer, indicating that the increased EDOs are also related to the stabilization by neighbor molecules. The calculated non-bonding energy and the statistical results indicate that the host-dopant interaction plays a pivotal role for orienting heteroleptic Ir complexes and the force applies to the alignment of the iridium-pyridine bonds of the phosphors toward horizontal direction.

The type and magnitude of the non-bonding interactions are different for different phosphors and hosts, leading to different EDOs. For instance, Ir(ppy)$_2$tmd and Ir(3′,5′,4-mppy)$_2$tmd have a quadrupole composed of two dipoles from pyridines (δ+ charge) to Ir atom (2δ– charge). If there is a dipole in host molecule (i.e., TSPO1), strong dipole and quadrupole interaction [–P=O$^{\delta-}$ and $^{\delta+}$H(pyridine)] anchors one phosphor molecule to two host molecules, leading to rather horizontal orientation of iridium-pyridines bond of the phosphors which is approximately parallel to the TDM. In contrast, if host molecule has positive surface potential [i.e., $^{\delta+}$(phenyl)$_3$-Si-phenyl-Si-(Phenyl)$_3^{\delta+}$ in UGH2], there must be repulsive force between pyridine of phosphors and host molecules so that pyridine ring must be pushed to vacuum. Dispersion force between the conjugated phenyl substituents of Ir(dmppy-ph)$_2$tmd and nearest neighbors anchors the pyridines onto the surface as well and lowers the energy with the molecular long axis lying on the surface. Meanwhile, random EDO of Ir(ppy)$_3$ is attributed to orthogonal intermolecular interaction sites, resulting in random orientation of the molecule.

Fig. 5a, 5b, and 5c exhibit the representative molecular behaviors of phosphors and nearest host molecules during the deposition out of 50 cases (Fig. S3 in Supporting Information) with different host-dopant combinations of UGH-2:Ir(ppy)$_2$tmd, TSPO1:Ir(ppy)$_2$tmd, and TSPO1:Ir(dmppy-ph)$_2$tmd, respectively. Large vibrations, rotations and diffusions of Ir(ppy)$_2$tmd on the surface of UGH-2 layer without lowering the energy were observed in the trajectory shown in Fig. 5a. The perpendicular alignment of pyridines occasionally formed on the surface resulted in vertical emitting dipole moment in average. On the other hand, a hydrogen atom at one of pyridines of Ir(ppy)$_2$tmd faced toward an oxygen atom of TSPO1 with the −P=O··H(pyridine) distance around 0.4 nm at t=2,472 ps and t=4,560 ps, thereby the parallel alignment of the Ir-pyridines of Ir(ppy)$_2$tmd to the surface. Larger quadrupole moment of Ir(3′,5′,4-mppy)$_2$tmd than that of Ir(ppy)$_2$tmd increased the strength of quadrupole-dipole interaction and resulted in enhancement of fraction of the horizontal dipole compared to Ir(ppy)$_2$tmd. The horizontal EDO of Ir(ppy)$_2$tmd in CBP could be understood by the dipole inducement in carbazole groups of CBP when the positive pole of pyridine approaches but the interaction strength for the iridium−pyridines alignment between Ir(ppy)$_2$tmd-CBP is smaller than that between Ir(ppy)$_2$tmd−TSPO1. Compared to the former cases, the picture of Ir(dmppy-ph)$_2$tmd shown in Fig. 5c is rather simple. The phosphor deposited onto the TSPO1 layer was stabilized after short time of deposition to the one with the horizontal iridium-pyridine-phenyl alignment and maintain the configuration. Much lower (larger) non-bonding energy of Ir(dmppy-ph)$_2$tmd in Fig. 4b restrained rotation of the molecule and the molecular configurations are easily fixed on the surface, resulting in much enhancement of horizontal EDO was achieved by the substitutions.

## Conclusion

We investigated the origin of the molecular orientation and EDO of doped heteroleptic iridium complexes in vacuum-deposited organic layers using MD simulations and QM analyses in direct comparison with experimental observation. Careful analyses on the simulation results revealed that molecular alignments of the phosphors are spontaneous by local electrostatic and van der Waals interaction with nearest host molecules interacting in a smaller scale than a molecule. Orientation of the TDM vector of the phosphors on the organic surfaces follows the direction of the ligand mainly participating in optical transition, such as pyridines in ppy, in the molecular alignment, whereas the alignment of ancillary ligand does not have a direct correlation with the EDO. Attractive interactions between pyridines of a phosphor and CBP (quadrupole–induced dipole interaction) or TSPO1 (quadrupole–dipole interaction) anchor the phosphor onto host molecules with the parallel iridium-pyridine alignment, thereby increasing the horizontal EDO. Ir(3′,5′,4-mppy)$_2$tmd has larger quadrupole moment than Ir(ppy)$_2$tmd resulted in further molecular alignment for the horizontal emitting dipole moment. Increase of the dispersion force along the direction of TDM was also effective on control of the molecular orientation for the horizontal EDO with lowered non-bonding energy.

## Methods

### *Quantum mechanical calculations and molecular dynamics simulations*

Density functional theory (DFT) was used to obtain molecular geometries and electrostatic potentials of host and phosphors. Triplet TDMs of the same phosphors from $T_1$ to $S_0$ were also calculated via SOC-TDDFT. The DFT and SOC-TDDFT calculations and the follow-up analyses were performed with Schrödinger Materials Science Suite[23] along with the quantum chemical engine, Jaguar.[27] A TDM having the largest oscillator strength among the three degenerated states of $T_1$ level ($T_x$, $T_y$, and $T_z$) obtained from the density functional calculations was used in this study as a representative transition dipole moment. All MD simulations were performed by Desmond,[28-29] a molecular dynamics engine implemented in the Schrodinger Materials Science Suite. Equilibration simulations prior to deposition were performed in NPT ensembles where pressure and temperature were set constant via Nose-Hoover chain and Martyna-Tobias-Klein method, respectively. There were no explicit constraints to geometry and/or positions to any of the molecules that were introduced in the simulation box. The simulations were performed over NVIDIA general-purpose GPU cards (K80).

*Rotation matrix method*

Rotation matrix method was used for transformation of the TDM vector from the molecular coordinate to the laboratory coordinate. Rotation angles of $\alpha$, $\beta$, and $\gamma$ are defined as the clockwise rotations to laboratories axes of $\boldsymbol{n}_x$, $\boldsymbol{n}_y$, and $\boldsymbol{n}_z$ axes, respectively. Then, the rotation matrixes for $\alpha$, $\beta$, and $\gamma$ rotations are followings:

$$R_\alpha = \begin{pmatrix} 1 & 0 & 0 \\ 0 & \cos\alpha & \sin\alpha \\ 0 & -\sin\alpha & \cos\alpha \end{pmatrix}, \tag{4}$$

$$R_\beta = \begin{pmatrix} \cos\beta & 0 & -\sin\beta \\ 0 & 1 & 0 \\ \sin\beta & 0 & \cos\beta \end{pmatrix}, \tag{5}$$

$$R_\gamma = \begin{pmatrix} \cos\gamma & \sin\gamma & 0 \\ -\sin\gamma & \cos\gamma & 0 \\ 0 & 0 & 1 \end{pmatrix}. \tag{6}$$

Sequential $\alpha\beta\gamma$ rotations of the dopant molecules were extracted in every configurations of the MD simulation. A product of the three rotation matrixes gives a matrix representing the orientation of the dopant molecule by

$$R_{\text{total}} = R_{\gamma} R_{\beta} R_{\alpha}. \tag{7}$$

Finally, the TDM vectors in the laboratory coordinate were obtained by

$$\text{TDM}_{\text{Lab}} = R_{\text{total}} \text{TDM}_{\text{Mol}}. \tag{8}$$

**Appendix**

*Calculation of a probability density function of* $TDM_H$.

Integration of a probability density function, *f*, indicates a probability of a variable *X* between *X* = *a* and *b*.

$$P(a < X < b) = \int_a^b f_X(x)dx, \tag{A1}$$

where

$$\int_{-\infty}^{\infty} f_X(x)dx = 1. \tag{A2}$$

To calculate the probability density function of $\sin^2 \varphi$ from an arbitrary vector, we define an arcsine function of

$$y = \arcsin(\sqrt{x}),\ \ 0 \le x \le 1, \tag{A3}$$

which is a reversed function of $y = \sin^2 x$. For a monotonic function, the variables are related by

$$f_X(x) = f_Y(y)\frac{dy}{dx}. \tag{A4}$$

If we put $f_Y(y) = \sin(y)$ and $\frac{dy}{dx} = \frac{1}{2\sqrt{x - x^2}}$ into equation (A4), the probability density function $(f_X)$ is obtained as

$$f_X(x) = \frac{1}{2\sqrt{1 - x}}. \tag{A5}$$

**Acknowledgment**

This work was supported by the Midcareer Research Program through an NRF (National Research Foundation) grant funded by the MSIP (Ministry of Science, ICT and Future Planning) (2014R1A2A1A01002030). We appreciate the helpful comments from Drs. Mathew D. Halls and H. Shaun Kwak from Schrödinger Inc., Prof. Youn Joon Jung from Seoul National University, and Dr. Denis Andrienko from Max Planck Institute for Polymer Research (MPIP).

**Author contributions**

C.-K.M. prepared samples, designed experiments, demonstrated calculations and simulations, prepared mathematical models, and prepared the manuscript; K.-H.K. prepared materials; J.-

J.K. directed experiments, calculation, and the manuscript.

**Additional information**

Supplementary information is available in the online version of the paper. Reprints and permissions information is available online at www.nature.com/reprints.

Correspondence and requests for materials should be addressed to H.T. or A.L.

**Competing financial interests**

The authors declare no competing financial interests.

## Figures

**Figure 1**. **Method for simulation of the EDO of emitters in vacuum-deposited layers.** (**a**) Transfer of the TDM vectors (red arrow) in the molecular coordinates to the vectors of the molecules on the organic substrate during the vacuum deposition simulation. (**b**) Three rotation angles ($\alpha$, $\beta$, and $\gamma$ for the clockwise rotation to the $\boldsymbol{n}_X$-, $\boldsymbol{n}_Y$-, and $\boldsymbol{n}_Z$-axes, respectively) were the orientation parameters of the molecules to correlate the molecular orientation to the laboratory axis. Angles between $\boldsymbol{n}_z$ axis and the TDM vector $(\varphi_L)$ and the $C_2$ axis $(\varphi_C)$ are obtained after the vector transformation. (**c**) A simulation box consisting of the substrate and target molecules. 50 target molecules were located above the substrate with 5.0 nm of inter-planar space dropped individually at 300 K. The distance unit in the figure is angstrom (Å).

**Figure 2. Iridium complexes and host materials.** (**a**) Chemical structures, transition dipole moment vectors, and electrostatic potentials of $Ir(ppy)_2tmd$, $Ir(3',5',4\text{-}mppy)_2tmd$, and $Ir(dmppy\text{-}ph)_2tmd$ phosphors. There are linear quadrupoles in the ground state of $Ir(ppy)_2tmd$ and $Ir(3',5',4\text{-}mppy)_2tmd$ with quadrupole moments along the principal axes of $Q_{xx,yy,zz}$ = [25.2,-13.0,-12.2] and [27.1,-12.8,-14.3] Debye·Å$^2$, respectively. (**b**) Chemical structures and electrostatic potentials of UGH-2, CBP, and TSPO1 host molecules. The electrostatic potentials are projected on the isosurface of electron density of 0.005 electrons/bohr$^3$. Optimization of the molecular structures were demonstrated using B3LYP method and LACVP** basis set for the phosphors and 6-31g(d)** for the host materials, respectively. SOC-TDDFT of the phosphors were carried out using B3LYP method and DYALL-2ZCVP_ZORA-J-PT-GEN basis set.

**Figure 3**. **Histograms of the EDO and angle of the $C_2$ axis of the phosphors in 5 host-**

**dopant combinations from the deposition simulation.** Each histogram includes 41,700 data in total from the configurations during 50 cases of the deposition in steps of 6 ps in the time regions of 1-6 ns. Data in the time region less than 1 ns were not used in the statistical analysis to exclude the steps of adsorption and the initial equilibration. (**a**) Histograms of the EDO with simulated Θ values. Red bars indicate population of the phosphor configurations having $TDM_H$ values in steps of 0.01. Blue lines are theoretical lines of $TDM_H$ from an arbitrary vector of which detailed derivation is given in Appendix. Green lines represent deviations of the population compared to the distribution of $TDM_H$ of an arbitrary vector. (Inset: enlarged deviation in the region of $0.8 \leq TDM_H \leq 1$) (**c**) Stacked histogram of the angle of the $C_2$ axis of phosphors and mean angles. Populations of the vector are plotted in steps of 2°. Distribution of the angle in different ranges of $TDM_H$ is distinguished by different colors.

**Figure 4. Schematic illustration of heteroleptic Ir-complexes having horizontal and vertical transition dipole moments and non-bonding energy of phosphors depending on the dipole orientation.** (**a**) Blue, gray, and red spheres at the octahedral sites represent pyridine rings, phenyl rings, and the ancillary ligand (–tmd), respectively. Dark blue and red arrows indicate the molecular $C_2$ axis and the TDM vector, respectively. Five configurations of the molecule for horizontal TDM and one configuration for vertical TDM are illustrated depending on the angle of the $C_2$ axis. (**b**) Calculated bon-bonding energy with cut-off radius of 0.9 nm of each atom of the phosphors as a function of $TDM_H$ in the five host-dopant systems.

**Figure. 5. Representative configurations and non-bonding energy of the phosphor on the surface during the deposition.** Snapshots of local configurations and time-dependent trajectories of the EDO, angle of the $C_2$ axis, and non-bonding energy up to 6 ns are depicted together. The ancillary ligand and pyridine rings of the phosphors at the octahedral sites are

colored by red and blue, respectively. (**a**) Ir(ppy)$_2$tmd deposited onto the UGH-2 layer has continuous rotation and the occasionally observed perpendicular alignment of pyridines with respect to the substrate results in vertical EDO. (**b**) Ir(ppy)$_2$tmd anchors on the surface of TSPO1 layer by local quadrupole-dipole interaction between the two nearest host molecules located at both sides. The hydrogen atoms at both pyridines of Ir(ppy)$_2$tmd and the oxygen atoms of TSPO1 connected by a broken line were the plausible binding sites. The distances between the two atoms (broken lines) are getting closer until around 0.4 nm as the time increases and a host-dopant-host pseudo-complex is formed with the parallel alignment of pyridines with respect to the substrate. (**c**) Ir(dmppy-ph)$_2$tmd deposited onto the TSPO1 layer are less mobile than Ir(ppy)$_2$tmd with the low non-bonding by the configuration of large dispersion force energy along the direction of TDM.

**Tables**

**Table 1. Comparison of simulated and measured EDOs in 5 combinations of host and heteroleptic Ir complexes in addition to Ir(ppy)$_3$ doped in the CBP layer for reference.**

**Fig. 1**

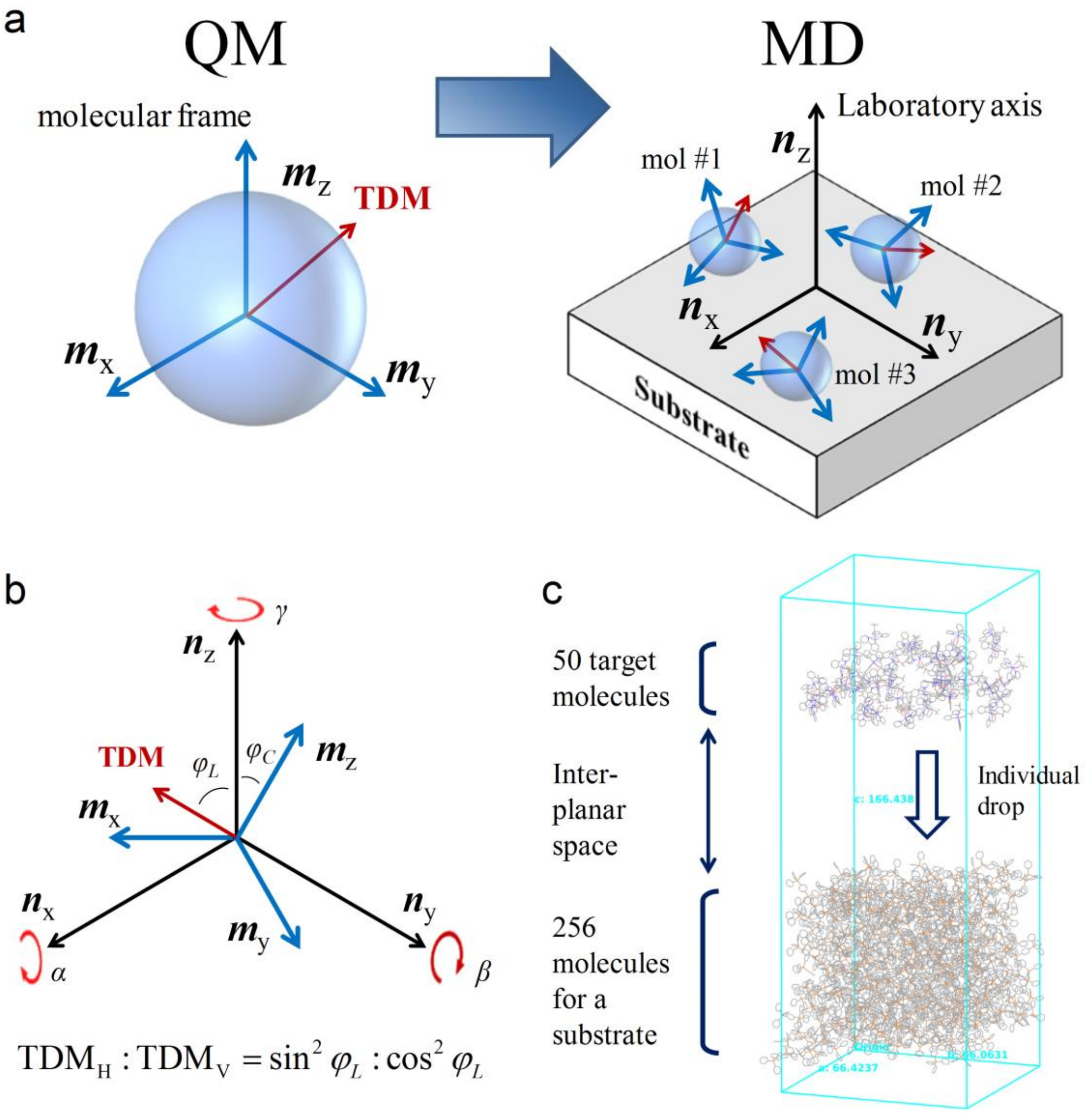

$$\mathrm{TDM_H} : \mathrm{TDM_V} = \sin^2 \varphi_L : \cos^2 \varphi_L$$

**Fig. 2**

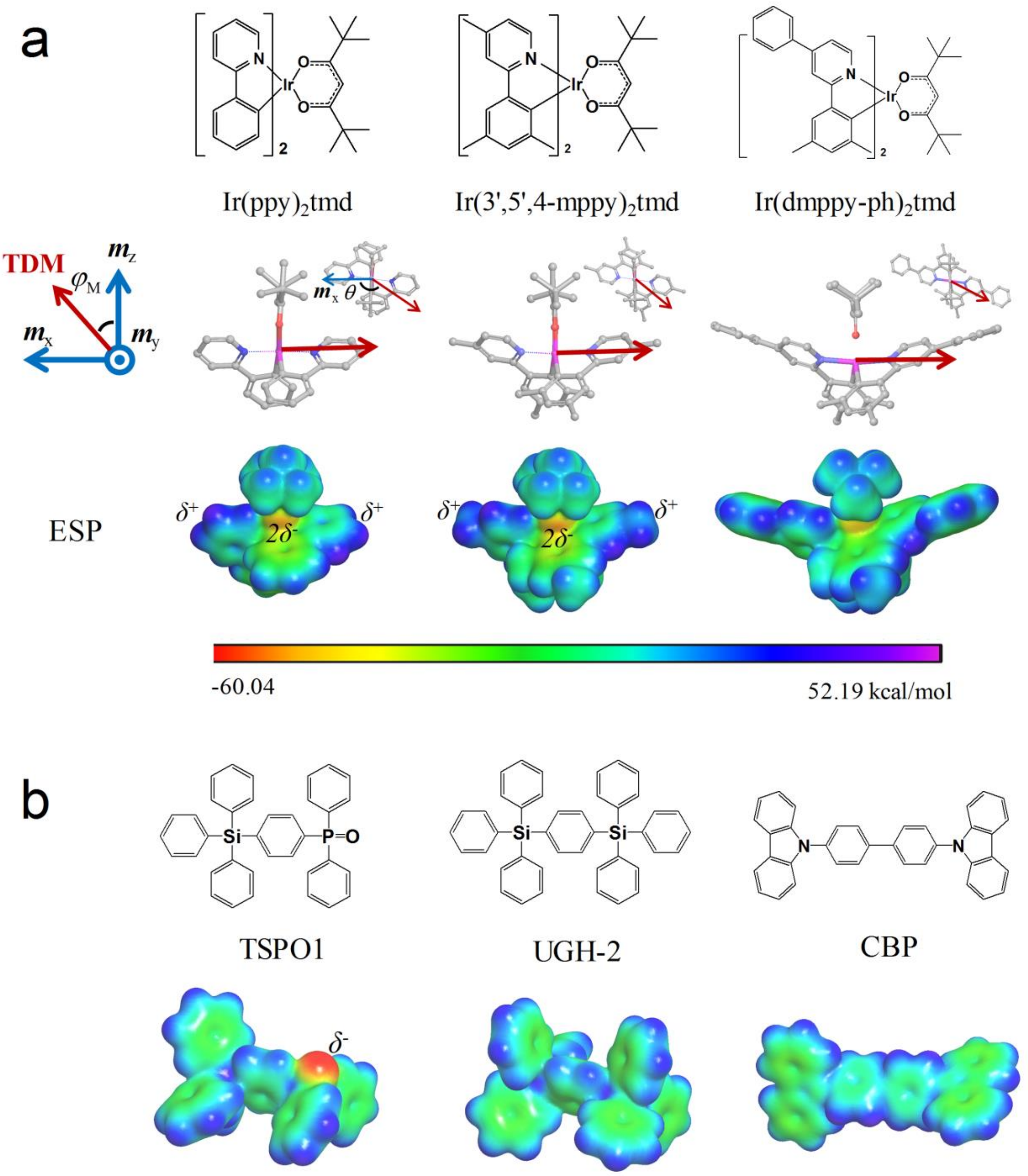

**Fig. 3**

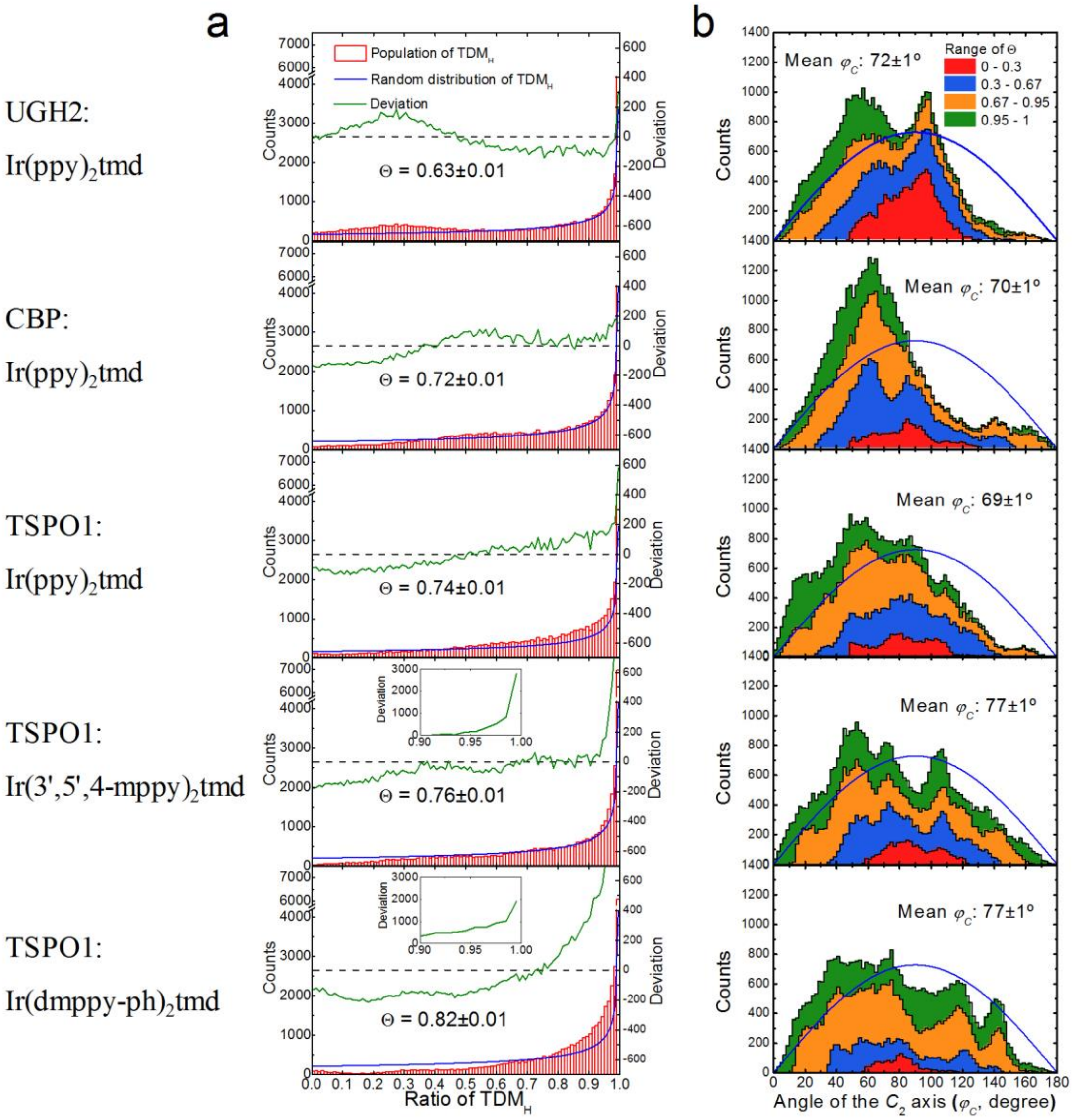

**Fig. 4**

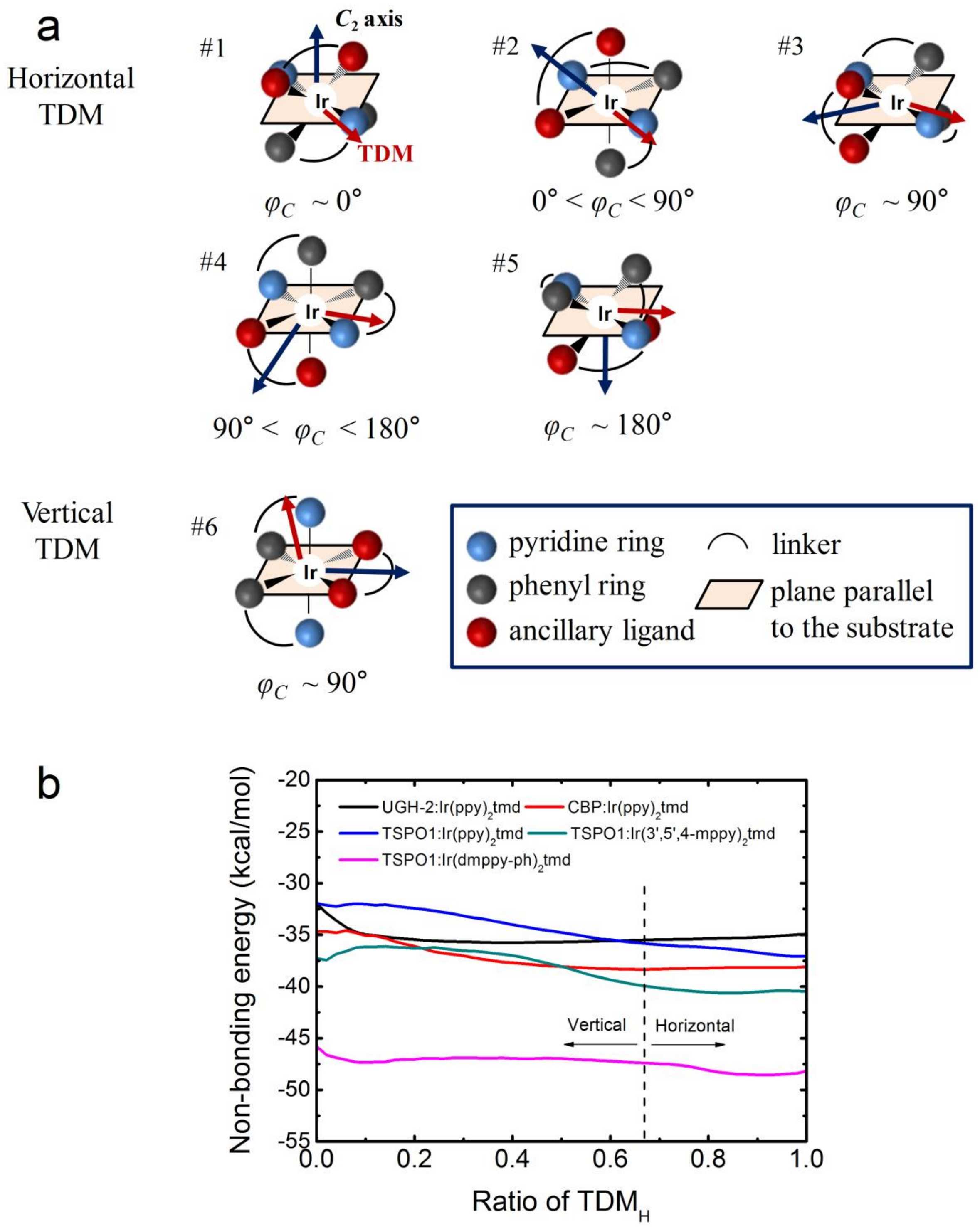

**Fig. 5**

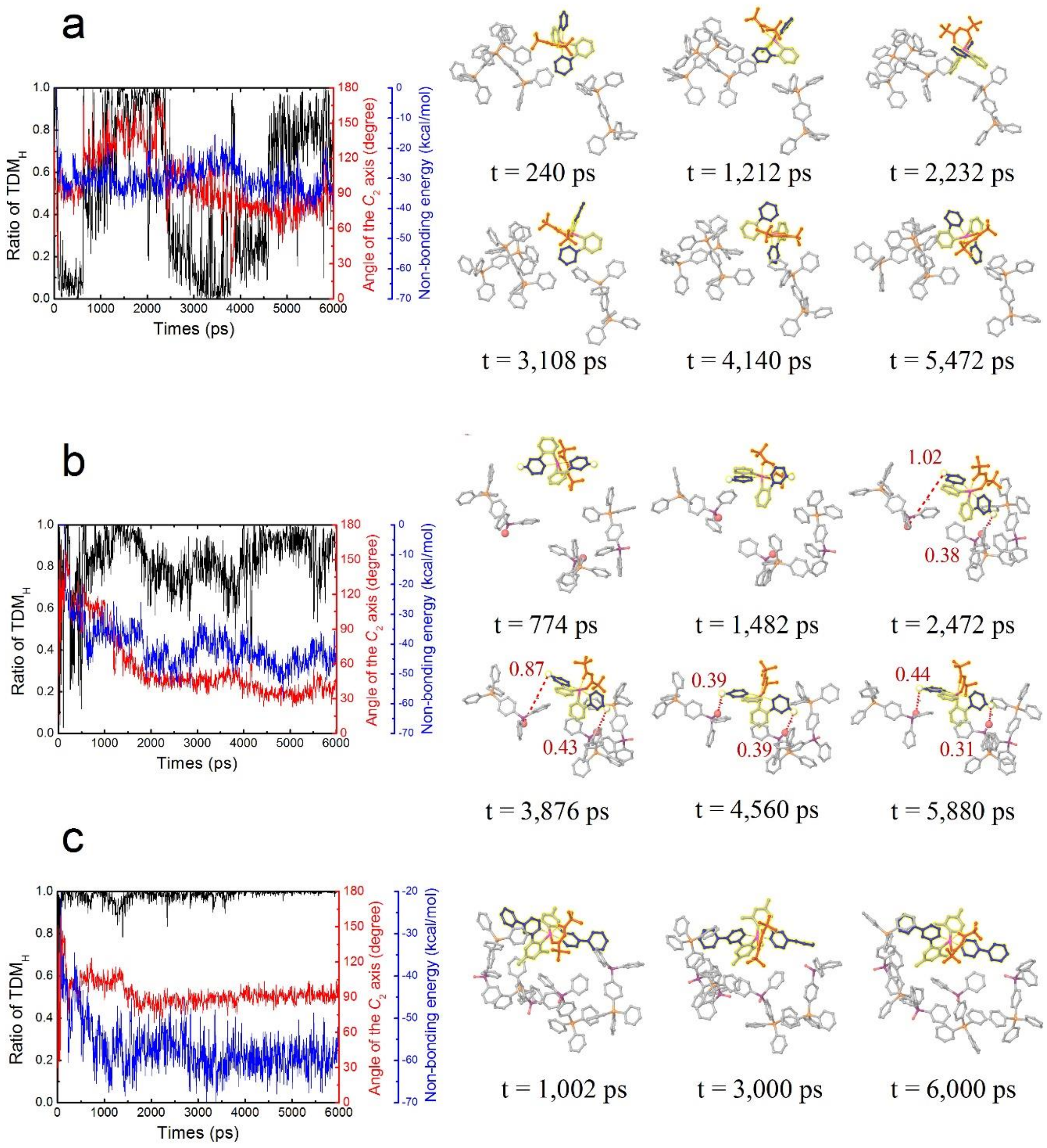
a
Ratio of TDM_H
Angle of the C2 axis (degree)
Non-bonding energy (kcal/mol)
Times (ps)
t = 240 ps
t = 1,212 ps
t = 2,232 ps
t = 3,108 ps
t = 4,140 ps
t = 5,472 ps
b
Ratio of TDM_H
Angle of the C2 axis (degree)
Non-bonding energy (kcal/mol)
Times (ps)
t = 774 ps
t = 1,482 ps
1.02
0.38
t = 2,472 ps
0.87
0.43
0.39
0.39
0.44
0.31
t = 3,876 ps
t = 4,560 ps
t = 5,880 ps
c
Ratio of TDM_H
Angle of the C2 axis (degree)
Non-bonding energy (kcal/mol)
Times (ps)
t = 1,002 ps
t = 3,000 ps
t = 6,000 ps

**Table. 1**

| Host | CBP | UGH-2 | CBP | TSPO1 | TSPO1 | TSPO1 |
|---|---|---|---|---|---|---|
| Dopant | $Ir(ppy)_3$ | $Ir(ppy)_2tmd$ | $Ir(ppy)_2tmd$ | $Ir(ppy)_2tmd$ | $Ir(3',5',4\text{-}mppy)_2tmd$ | $Ir(dmppy\text{-}ph)_2tmd$ |
| Simulation | 67:33 | 63:37 | 72:28 | 73:27 | 76:24 | 82:18 |
| Measurement | 67:33 | 60:40 | 75:25 | 78:22 | 80:20 | 86:14 |

**Supplementary Information**

*Unraveling the orientation of phosphors doped in organic semiconducting layers*

Chang-Ki Moon, Kwon-Hyeon Kim, and Jang-Joo Kim*

Department of Materials Science and Engineering, RIAM, Seoul National University, Seoul 151-744, South Korea.

*Preparation of organic substrates by MD*

Preparation of the substrate had three steps of the MD simulation after locating 256 molecules in a grid: annealing at 500K at first (NVT, 500 ps) followed by annealing at 300 K (NVT, 200 ps), and finally packing of the molecules at 300K and 1 atm (NPT, 1,000 ps for UGH-2 and CBP, and 5,000 ps for TSPO1). Fig. S1a shows the trajectories of the density and total potential energy of UGH-2, CBP, and TSPO1 substrates, respectively, at the step of packing. Density and potential energies were converged during the 1,000 ps of simulation for the UGH-2 and CBP substrates and 5,000 ps of simulation for the TSPO1 substrate. The NPT MD simulation produced amorphous solid densities of 1.04, 1.13, and 1.10 g/cm$^3$ for UGH-2, CBP, and TSPO1, respectively. Consider for example CBP whose simulated density of 1.13 g/cm$^3$ is in good agreement with the experimental value of 1.18 g/cm$^3$.[S1] Fig. S1b exhibits the orientation of 256 molecular vectors of UGH-2, CBP, and TSPO1 indicated as red arrows, respectively, consisting the substrates. Their angular distributions were closed to the random distribution line so we concluded that the amorphous organic substrates were successfully prepared.

*Film fabrication and angle-dependent PL measurement*

Organic films were fabricated on fused silica substrates by thermal evaporation under a pressure of $5\times10^{-7}$ torr without breaking the vacuum. Rates of the co-deposition were 1 Å/s and thickness of the films were 30 nm. Films were encapsulated using glasses and UV resin after the deposition. EDOs of the organic films were measured by angle-dependent PL.[18] The fused silica substrates were attached to a half-cylinder lens made of fused silica with an index similar densities and matching oil. The attached films were excited by a He-Cd laser (325 nm, CW) and the angular PL spectra escaping out through the lens were measured by Maya2000 spectrometer (Ocean Optics Inc.). A linear polarizer was used to obtain the *p*-polarized PL patterns. Optical simulation of the luminescence from a thin film was employed to fit the dipole orientation from the PL patterns.[S2]

Fig. S2a shows the angular *p*-polarized emission patterns at peak wavelengths of $Ir(ppy)_2tmd$, Ir(3′,5′,4-mppy), and $Ir(dmppy\text{-}ph)_2tmd$ doped in 30-nm-thick TSPO1 layers, respectively. Lines are the calculated emission patterns of the *p*-polarized light with an isotropic (blue), fully horizontal (red), and the best-fitted EDO (black). The lower emission intensities of the dyes above 50° represent that the emitter has larger portion of the horizontal emitting dipole moment. As a result, EDO of $Ir(ppy)_2tmd$, Ir(3′,5′,4-mppy), and $Ir(dmppy\text{-}ph)_2tmd$ were determined as $\Theta$ = 0.78, 0.80, and 0.86, respectively. The orientation values fitted in all the spectral range of the phosphors as shown in Fig. S2b.

*Data of $TDM_H$ in 50 cases of the deposition simulation*

Raw data of $TDM_H$ in 50 cases of the deposition simulation in five host-dopant combinations are depicted in Fig. S3a to S3e. Data were arranged in the order of standard deviation between 1000 ns and 6000 ns and the numbers in the figure is the simulation numbers. The circled trajectories in UGH-2:Ir(ppy)$_2$tmd, TSPO1:Ir(ppy)$_2$tmd, TSPO1:Ir(dmppy-ph)$_2$tmd were used as representatives in Fig. 5a, 5b, and 5c, respectively.

*Deposition simulation of Ir(ppy)$_3$ on CBP substrate*

Deposition simulation of Ir(ppy)$_3$ was demonstrated on the pre-organized CBP substrate layer. Ir(ppy)$_3$ is a well-known homoleptic complex exhibiting isotropic emitting dipole orientation (EDO) as doped in organic host layers.[S3-S4] The identical QM and MD methods described in the main text were employed to analyze the orientation of Ir(ppy)$_3$ deposited on CBP layer using Jaguar[S5] and Desmond.[S6] The symmetric structure of Ir(ppy)$_3$ results in three equivalent triplet transition dipole moments (TDMs) by $^3$MLCT with polar coordinates of [$\varphi_M$ = 87°, $\theta$ = 25°], [$\varphi_M$ = 87°, $\theta$ = 145°], and [$\varphi_M$ = 87°, $\theta$ = -95°], with respect to the molecular $C_3$ axis as shown in Figure S4a. Molecular configurations and three TDM vectors of Ir(ppy)$_3$ in the deposition simulation were recorded every 6 ps until reaching 6,000 ps. Figure S4b and S4c show the statistical results of the ratio of horizontal TMD (TDM$_H$) and the angle of the $C_3$ axis of Ir(ppy)$_3$. The distribution of TDM$_H$ follows the random distribution line (blue line in Figure S4b) and the ensemble average of Θ=0.67 is in good agreement with the EDO observed by experiments. The angular distribution of the $C_3$ axis is close to the random distribution line (blue line in Figure S4c) as well.

*Non-bonding energies for different host-dopant systems*

Non-bonding energies of the phosphors located on the surface were calculated by the summation of van der Waals and Coulomb interaction energies with cut-off radius of 0.9 nm of each atom of the phosphors. Figure S5 shows the raw data of the calculated non-bonding energies of the phosphors in every frame of MD simulations (41700 scatters for a system) with their dipole orientations in the configuration. Mean non-bonding energies of the five systems are indicated as red lines, respectively, and are compared in Figure 4b in the main text to discuss the relationship between molecular orientation of phosphors and intermolecular interactions. Note that mean non-bonding energies are lowered as $TDM_H$ increase for $Ir(ppy)_2tmd$:CBP, $Ir(ppy)_2tmd$:TSPO1, Ir(3′,5′,4-mppy)$_2$tmd:TSPO1, and Ir(dmppy-ph)$_2$tmd:CBP but there is a broad energy trap in the vertical orientation region of $TDM_H$=0.1−0.5 for $Ir(ppy)_2tmd$:UGH-2.

*Movie of vacuum deposition simulation*

Supplementary videos were attached for showing an example of the vacuum deposition simulation. (The 7th deposition picture on 3rd line in Figure S3c, side view and top view on video 1 and 2, respectively). The ancillary ligand and pyridine rings of the phosphors at the octahedral sites are colored by red and blue, respectively, on the videos.

**Figure S1. Potential energy, volume, and molecular distribution of the organic substrates prepared by MD simulation.** (**a**) Change of the density and total potential energy of the substrates consisting of 256-number of UGH-2, CBP, and TSPO1 molecules, respectively, in the simulation of packing the molecules at 300 K and 1 atm. (**b**) Angular distributions of the 256-number of molecular vectors (red arrows) in the substrates. Their distribution followed the random angular distribution line, indicating the amorphous substrates were formed by the MD simulation.

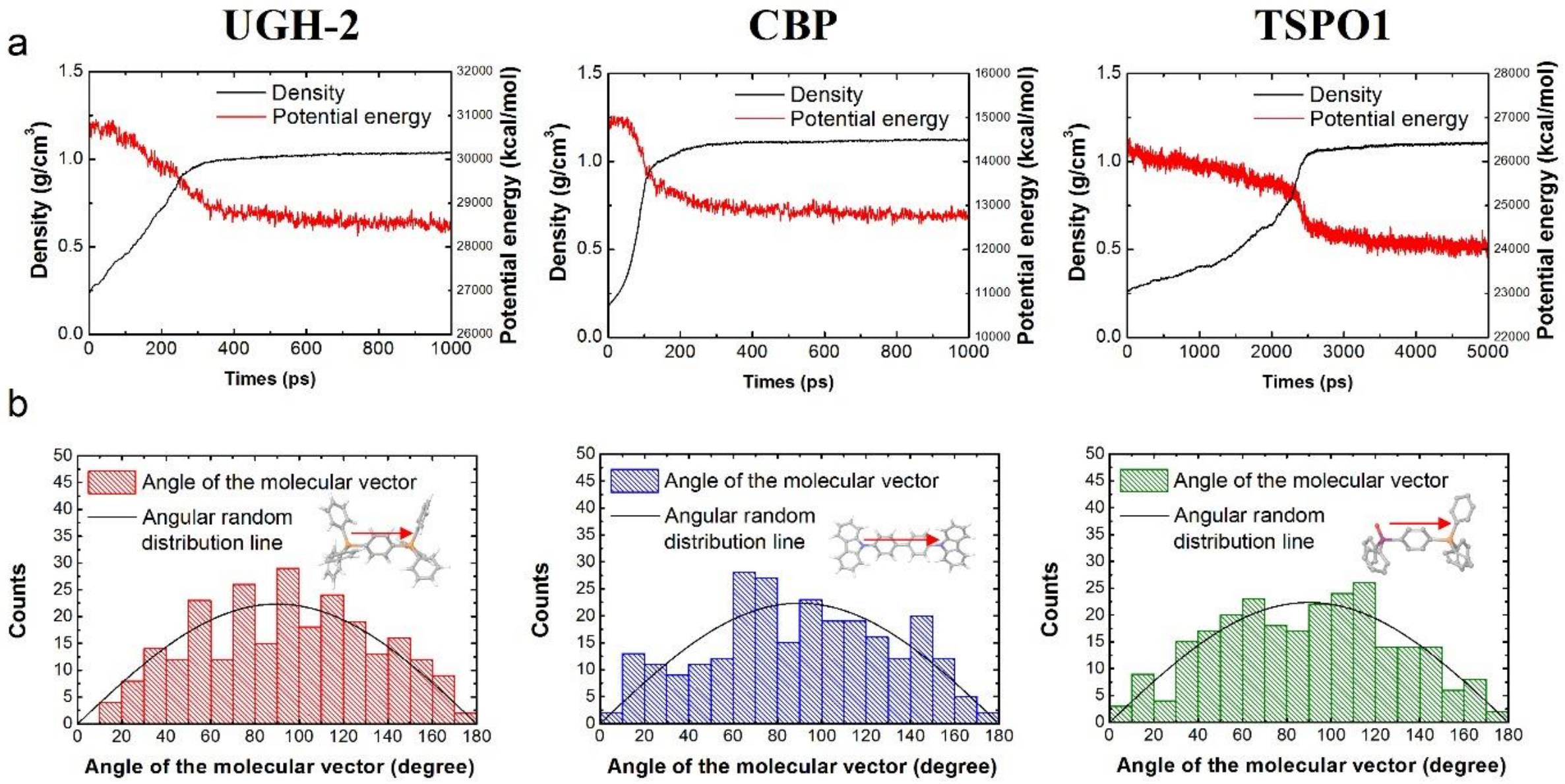

**Figure S2. Angle-dependent PL analysis of Ir(ppy)2tmd, Ir(3′,5′,4-mppy), and Ir(dmppy-ph)$_2$tmd doped in 30 nm of TSPO1 layers.** (**a**) Analysis of the EDO at peak wavelength of each emitter (520 nm, 530 nm, and 570 nm). Measured emission patterns (scatters) are located between the lines with an isotropic orientation (blue) and a fully horizontal orientation (red). The best fitted fractions of the horizontal to vertical emitting dipole moment were 78:22, 79:21, and 85:15, for Ir(ppy)$_2$tmd, Ir(3′,5′,4-mppy), and Ir(dmppy-ph)$_2$tmd, respectively. (**b**) Measured (surfaces) and calculated angular emission spectra (broken lines) having the orientation that have been determined by the analyses at the peak wavelengths.

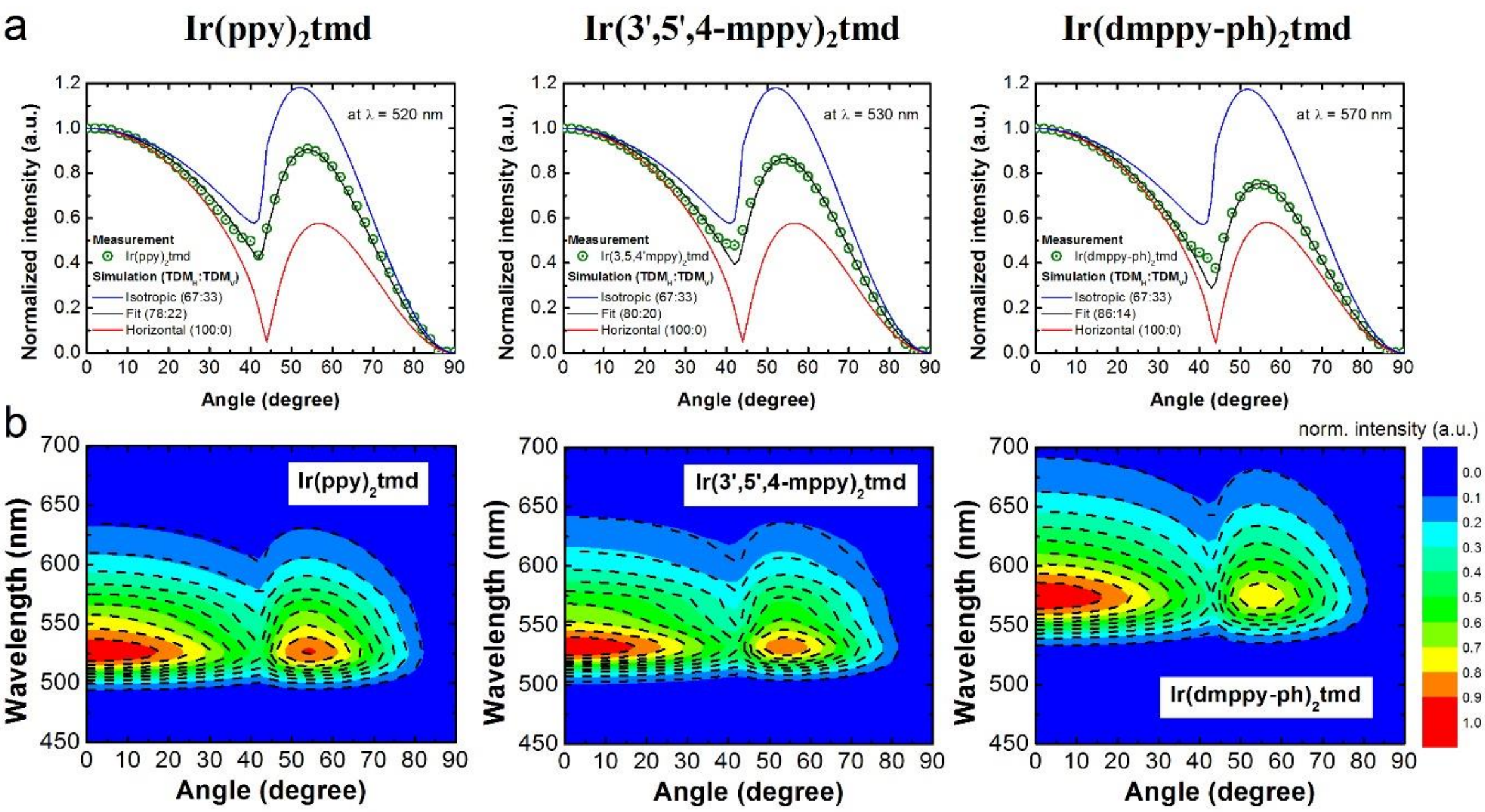

**Figure. S3. 50 trajectories of $TDM_H$ up to 6000 ps in 5 combinations of the host and the dopant.** Black and red lines represent the ratio of $TDM_H$ in the range of 0 to 1 and the angle of the $C_2$ axis of phosphors in the range of 0° to 180°, respectively.

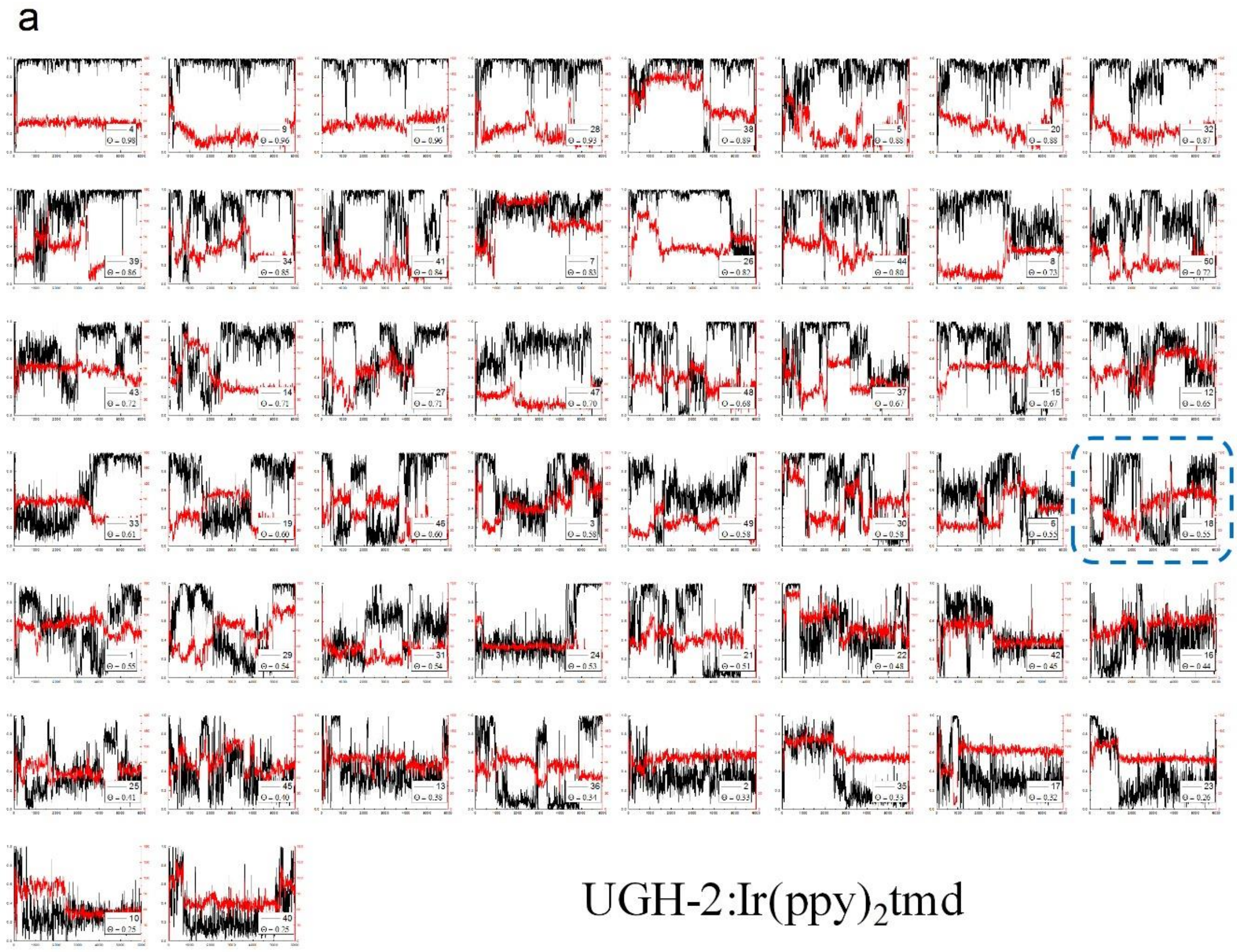

b

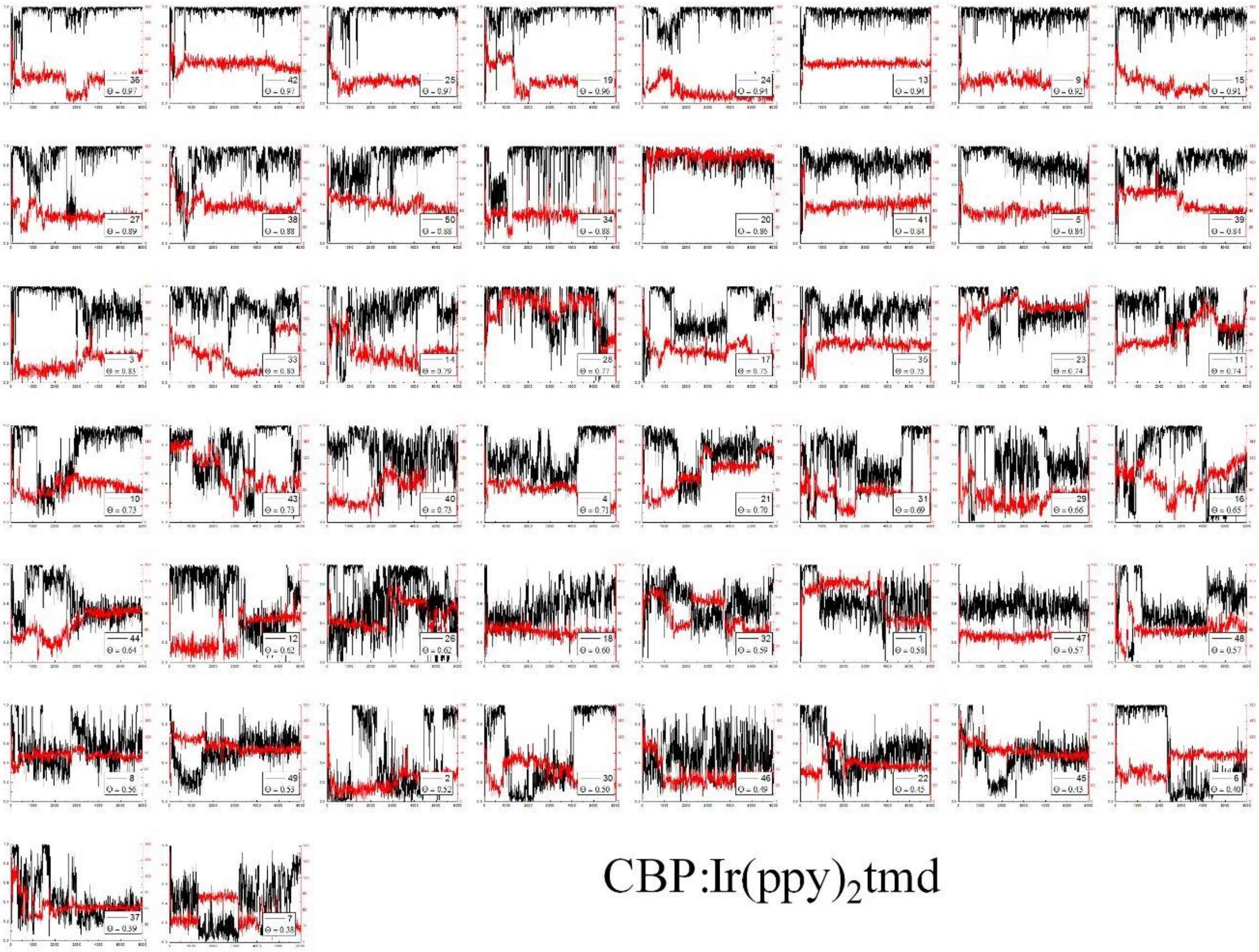

CBP:Ir$(ppy)_2$tmd

c

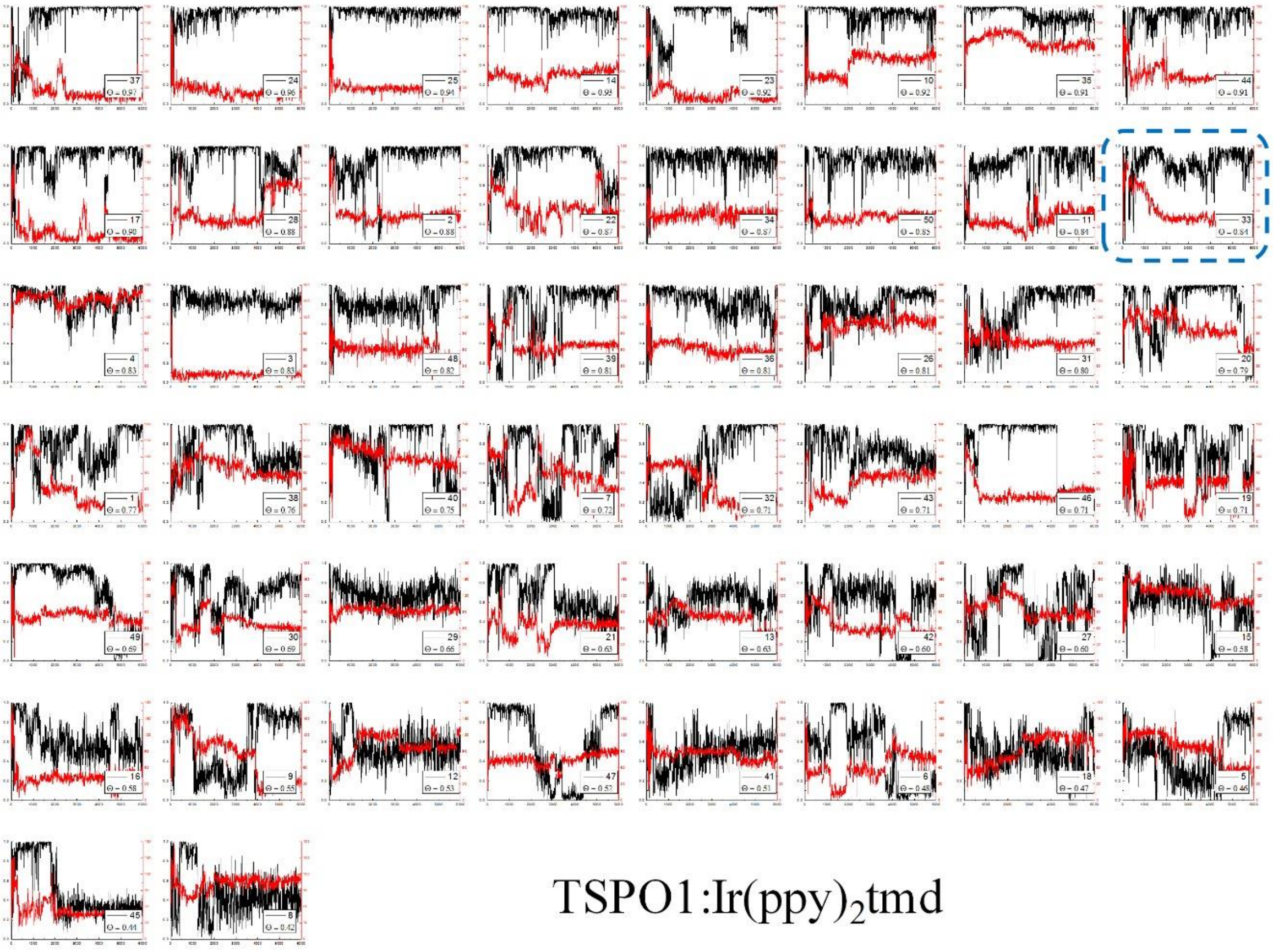

TSPO1:Ir(ppy)$_2$tmd

d

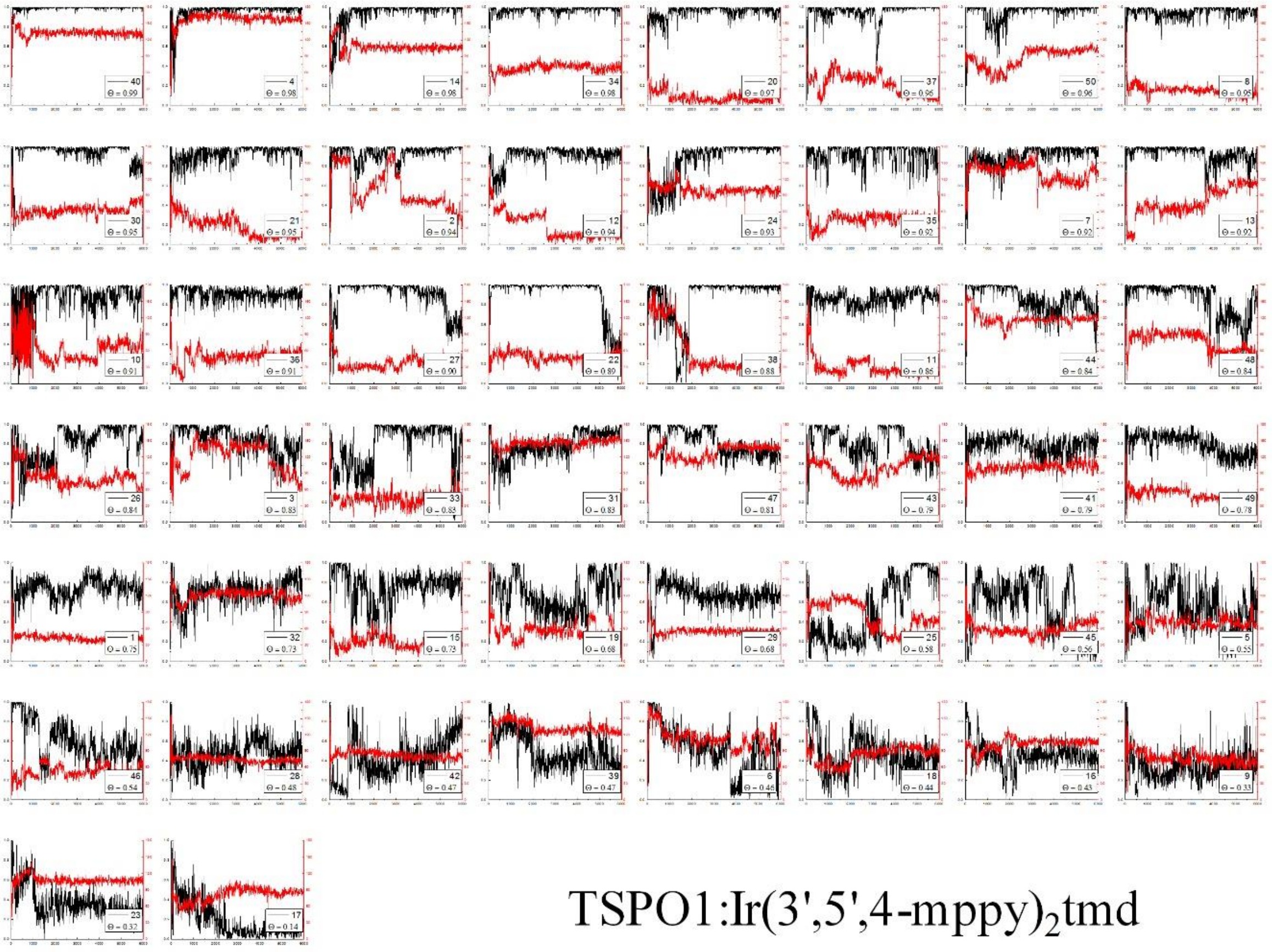

TSPO1:Ir(3',5',4-mppy)$_2$tmd

e

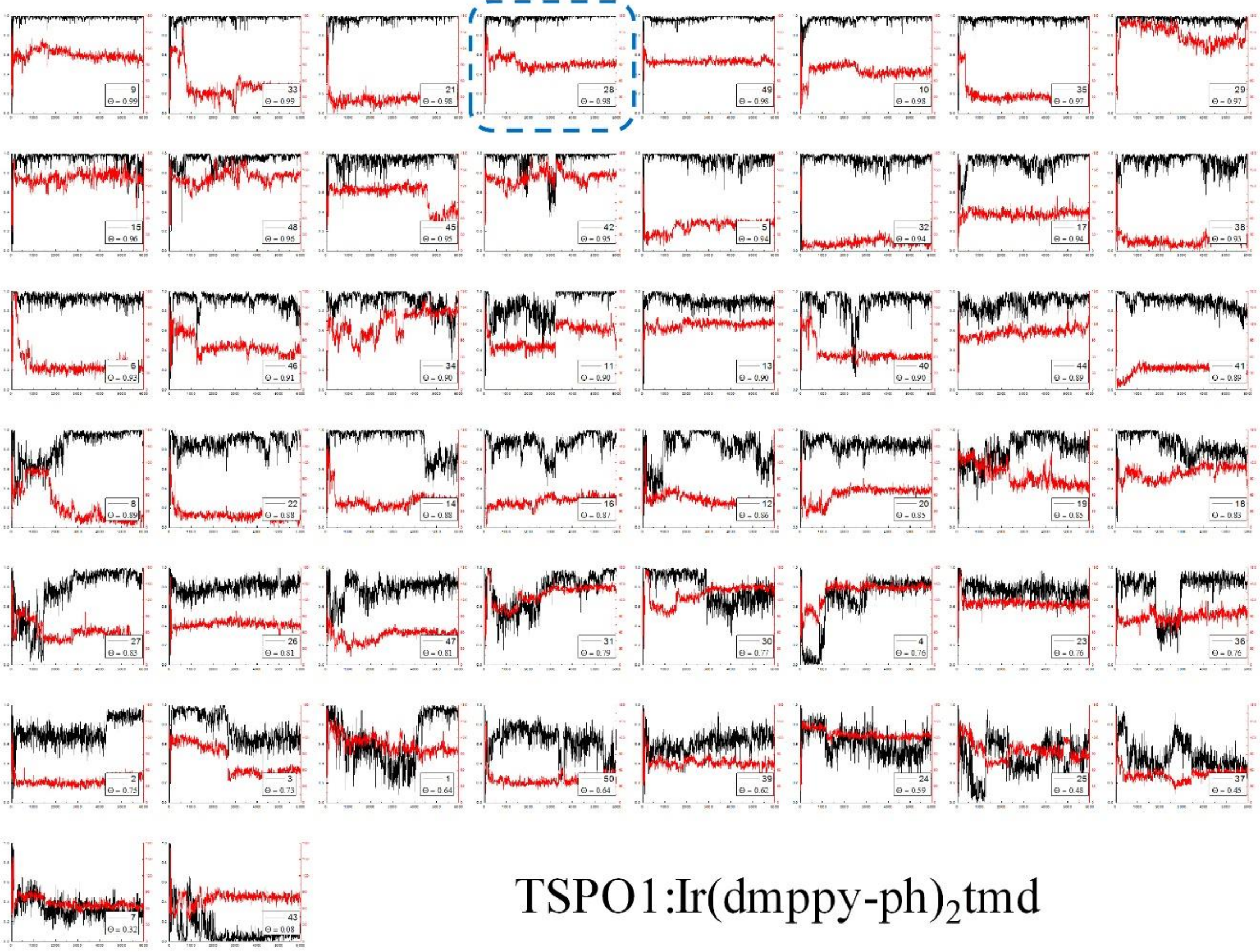

TSPO1:Ir(dmppy-ph)$_2$tmd

**Figure S4. Quantum chemical simulation of the TDMs and molecular dynamic simulation of molecular and emitting dipole orientations of $Ir(ppy)_3$.** (a) Three triplet TDM vectors of $Ir(ppy)_3$ with a 3-fold rotation symmetry from iridium to three equivalent ppy ligands by $^3$MLCT. The $C_3$ symmetry axis toward pyridines from the origin located at the Ir atom was set as $\boldsymbol{m}_z$, the vector normal to the plane including $\boldsymbol{m}_z$ and one of Ir-N vector was set as $\boldsymbol{m}_y$, and $\boldsymbol{m}_x$ was determined by a cross product of $\boldsymbol{m}_y$ and $\boldsymbol{m}_z$ in the dopants. Optimization of the molecular structures were demonstrated using B3LYP method and LACVP** basis set. Spin-orbit coupled time-dependent density functional theory (SOC-TDDFT) calculations were carried out using B3LYP method and DYALL-2ZCVP_ZORA-J-PT-GEN basis set. (b) A histogram of the $TDM_H$ of $Ir(ppy)_3$ with a simulated Θ value. Red bars indicate the population of the phosphor configurations having $TDM_H$ values in steps of 0.01. The blue line is the theoretical line of $TDM_H$ from an arbitrary vector and the green line represents the deviations between red bars and blue lines. Note that 125100 data were included in the histogram by a product of 41,700 frames and three TDMs. (c) A histogram of the angle of the $C_3$ axis of $Ir(ppy)_3$ in steps of 2°. The blue line represents an angular random distribution of an arbitrary vector. This histogram includes 41,700 data in total.

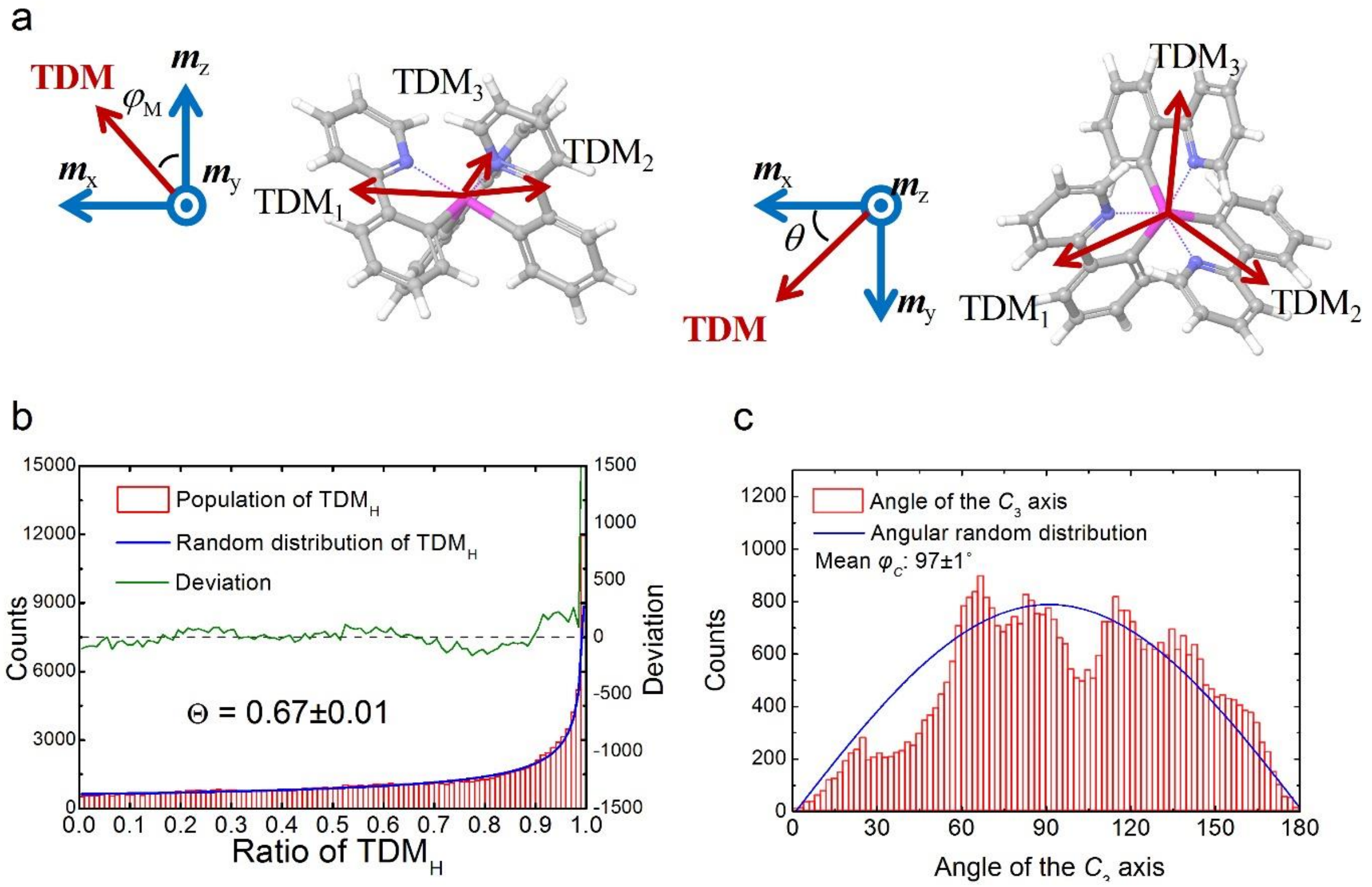

**Figure S5. Relationship of the emitting dipole orientation in five host-dopant systems**. Red lines represent the mean non-bonding energies as a function of the ratio of $TDM_H$.

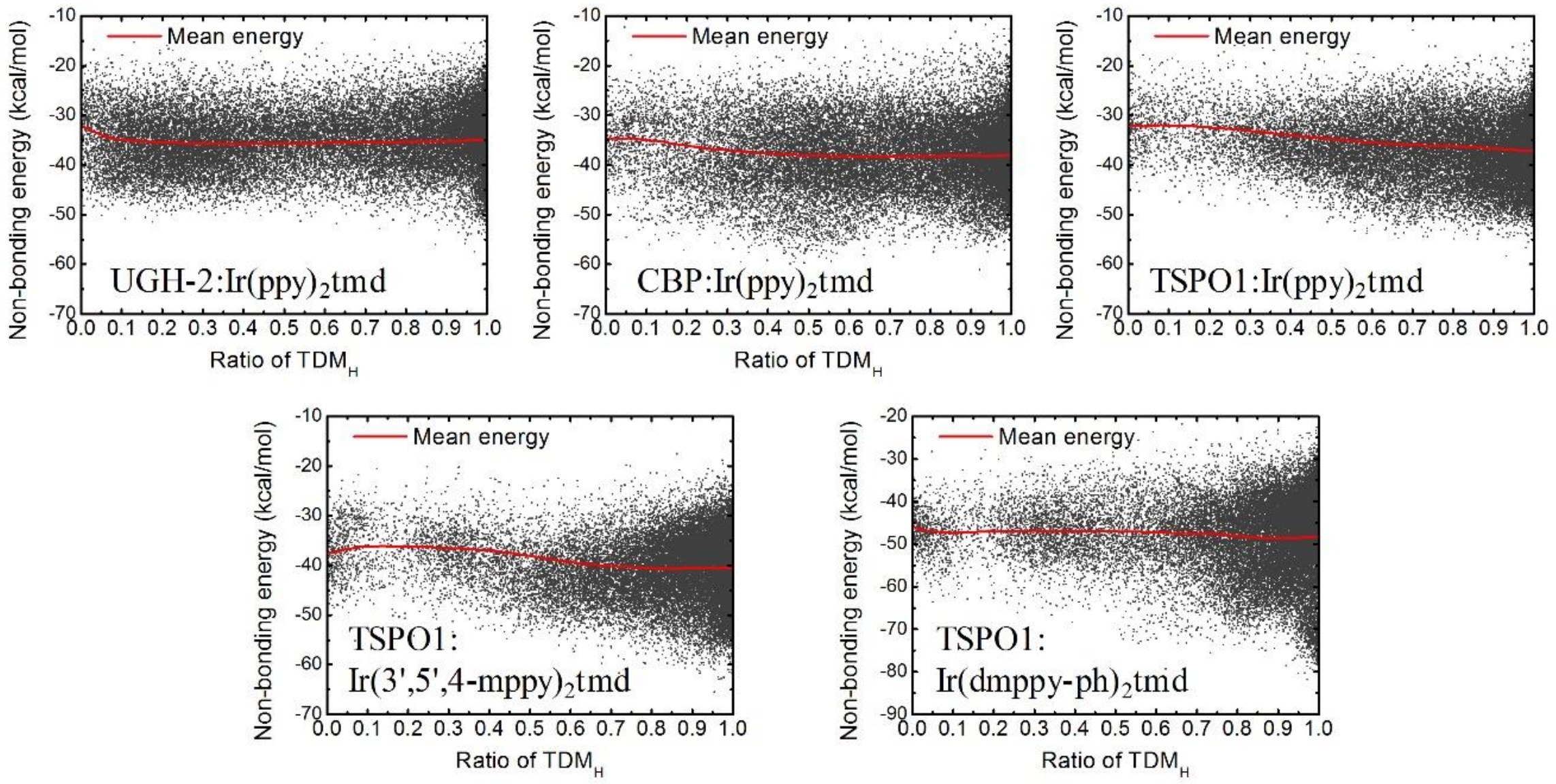